\documentclass[a4paper,preprint]{revtex4-1}
\usepackage{amsmath,amssymb,amsfonts,latexsym,graphicx}
\usepackage{indentfirst}

\newcommand{\concen}{C}
\begin{document}
\title{Is there a third order phase transition for supercritical fluids?}
\author{Jinglong Zhu}
\affiliation{LMAM and School of Mathematical Sciences,
Peking University, Beijing, P.R. China}
\affiliation{Beijing International Center for Mathematical Research, Beijing, P.R. China}
\author{Pingwen Zhang}
\affiliation{LMAM and School of Mathematical Sciences,
Peking University, Beijing, P.R. China}
\author{Han Wang}
\email{han.wang@fu-berlin.de}
\author{Luigi Delle Site}
\email{dellesite@fu-berlin.de}
\affiliation{Institute for Mathematics, Freie Universit\"at Berlin, Germany}

\begin{abstract}
  We prove that according to Molecular Dynamics (MD) simulations of
  liquid mixtures of Lennard-Jones (L-J) particles, there is no third
  order phase transition in the supercritical regime beyond Andrew's
  critical point. This result is in open contrast with recent
  theoretical studies and experiments which instead suggest not only
  its existence but also its universality regarding the chemical
  nature of the fluid. We argue that our results are solid enough to
  go beyond the limitations of MD and the generic character of L-J
  models, thus suggesting a rather smooth liquid-vapor thermodynamic
  behavior of fluids in supercritical regime.
\end{abstract}

\maketitle

{\it Introduction}:
When pressure and temperature increase beyond Andrew's
critical point~\cite{andrews1875, ma2011third} a fluid enters in the so-called
supercritical regime.  
Supercritical fluids are particularly
convenient as solvents in a wide variety of applications, for example,
for production of pharmaceutical powders~\cite{york2005supercritical}. However, despite their
existence is known for more than a century,
its potentiality started to be explored only in the last decades.
In this context, the experimental~\cite{nishikawa1995correlation,
  nishikawa1998fluid, morita2000study, nishikawa2000inhomogeneity, nishikawa2003density, arai2005analysis, sato2008structural,
  koga2009high} as well as theoretical~\cite{lotfi1992vapour, smit1992phase, panagiotopoulos1994molecular, potoff1998critical,
  perez2006critical, ma2011third} investigation of the fluid's behavior around
the critical point represents a mandatory task in order to clarify
the essential features of superfluidity and explore its
potentiality. In recent years the work of Yoshikata Koga and coworkers
claims the existence of the so-called ``Koga-Line'' in supercritical
regime, that is a collection of foci of anomalies in some third-order
derivatives of the Gibbs function~\cite{koga2009high}. The natural
consequence is the prediction of a third-order phase transition beyond
Andrews critical point; this latter, later on, was claimed in a
theoretical work by Ma and Wang~\cite{ma2011third}. In ref.\cite{ma2011third},
using a mean-field approach, not only is proved the existence of such a
transition but it is also claimed its universality regardless of the
specific molecular chemical structure. If such conclusions can be
proved true, then the physics of superfluidity will become by far more
clear: it would imply that by external manipulations of temperature
and pressure, thermodynamics quantities related to the second order
derivatives of Gibbs free energy, for example the heat capacity
$c_{P}$ or the isothermal compressibility $\kappa$, will have abrupt
variations which in turn may be programmed to change physical
properties, e.g. of solutes, on demand.  However the interpretation of
experimental results lies on the analysis of data according to ideal
models whose constraints may not be fully met by the experimental
conditions. On the other hand, theoretical models based on mean-field
approaches cannot properly characterize the instantaneous fluctuations of the particle number density
which are instead a key characteristic of fluids and whose
accurate description is mandatory in order to predict the correct behavior of second order derivatives
of the Gibbs free energy, among which, for example, the isothermal compressibility. In this
perspective it becomes mandatory to describe a fluid as a
particle based liquid and as a consequence the treatment of the
problem via molecular simulation. Unfortunately at the current state
of the art, in general, computationally affordable chemically detailed models 
for the supercritical regime are scarce~\cite{sakuma2013prediction, orabi2013polarizable} and even for those few available 
the capability of describing phase transitions is highly questionable (this is true even in standard thermodynamic
conditions) ~\cite{vega2011simulating}. However, generic Lennard-Jones liquids may be sufficient
for a satisfactory description of the mean features of the supercritical
state. The fact that a supercritical fluid has both liquid and gas
behavior suggests that the specific chemical structure and its
consequent bonding network are not as relevant as in ambient
conditions and thus a generic spherical (L-J) model may be able to
capture the essential thermodynamic features; 
this is a point that we will treat more specifically in the light of our results later on. In general, simulations of
L-J fluids have been for long employed to understand the thermodynamic
behavior around the critical point~\cite{panagiotopoulos1994molecular, perez2006critical},
however the exciting question of the possibility of the existence of a
third-order phase transition has never been addressed. Instead in a
previous paper, some of the authors of the current work have addressed
explicitly this question with extended numerical simulations and it
was proven that for a one-component L-J fluid a third-order phase
transition does not occur~\cite{wang2011existence}. Unfortunately this answer cannot
be considered satisfactory because the experimental conditions are
rather different and imply the use of (at least) a two-component fluid~\cite{koga2009high}.
In this perspective, here we have treated a much general
condition, that is we considered (I) a L-J fluid solvating L-J particles
of larger size, to mimic a situation of solvation, and (II) a binary mixture
of L-J particles with similar molecular size. Moreover since the critical point occurs at high
pressure where the interaction between L-J particles of different
nature becomes more relevant, we tested our conclusions also for the case of different
interaction strengths.  We will show that, as for the case of a
one-component liquid, there is a rather clear evidence that a third-order
phase transition does not occur.

{\it Model systems:}
In the present work, we consider 
particles interacting via the standard L-J potential.
The conventional notations are adopted: the interaction strength
and molecular diameter are denoted by $\epsilon$ and $\sigma$, respectively.
The one-component system has been already treated in our previous work~\cite{wang2011existence} and thus it will be considered only as a term of comparison here, while we extend the simulation study to systems (I) and (II).
We denoted the component of the mixture by molecules of type A and
type B,
so the interaction strength and molecular diameter
are denoted  $\epsilon_A$ and $\epsilon_B$,   $\sigma_A$ and $\sigma_B$,
respectively.
Molecules of different species also interact via the L-J potential, following the Lorentz-Berthelot
combination rule: $\epsilon_{AB} = \sqrt{\epsilon_A\epsilon_B}$ and $\sigma_{AB}=\frac{1}{2} (\sigma_A+\sigma_B)$. For Convenience, without loss of generality,
we assume $\sigma_A \leq \sigma_B$. The parameters that control the physics of the system, beside the thermodynamic parameters, are
the ratio of the interaction strength $E_{AB} = \epsilon_B / \epsilon_A$,
the ratio of the molecular diameter $R_{AB} = \sigma_B / \sigma_A$,
and the relative number density of B (concentration of B), denoted by
$\concen_B=N_B/ (N_A+N_B)$, where
$N_A$ and $N_B$ denote the number of molecules of specie A and specie B, respectively.
Our starting point is the one-component fluid, for which we have already proved the non-existence of the third order phase transition.
We choose this system as a reference and consider (build) system (I) and (II) as increasingly larger perturbations of the one-component system  (see details in Tab.~\ref{tab1}).
For simplicity, throughout this work,
we explicitly discuss  the case $E_{AB} = 1$. Several tests were carried out with different values of $E_{AB}$,
but no difference in the main conclusions was found, therefore the results are not presented here.
For (I), we initially insert a small number of large L-J solutes and then increase its concentration.
For (II), we mix first the reference one-component system with a fluid of molecules with very similar diameter, that is $R_{AB}=1.01$, at equal density concentration (i.e.~$C_B = 0.5$), and then progressively increase $R_{AB}$.
{The considered $R_{AB}$ ranges from $1.01$ to $1.05$, which means that we explore the
  perturbations to the case of a one component liquid in terms of
  molecular size of part of the system. These combined with different
  interaction strengths and solute concentrations will provide a
  description of how, going away from a standard one component liquid,
  the system reacts. If mixtures are more likely to go through a third
  order phase transition, a trend must clearly emerge.
}

\begin{table}
  \centering
  \caption{The tested
    controlling parameters of system I and II. $E_{AB}$ and  $R_{AB}$ are the ratios
    of the interaction strength and diameter of molecule type B over those of type A, respectively.
    $\concen_B$ is the number density of type B, defined by $\concen_B=N_B/(N_A+N_B)$.}
  \label{tab1}
  \begin{tabular*}{0.60\textwidth}{c@{\extracolsep{\fill}}cccc}
    \hline
    & $E_{AB}$ & $R_{AB}$ & $\concen_B$ \\      \hline
    System I  &1.0 & 2.00 & 0.002, 0.010, 0.020 \\
    System II &1.0 & 1.01 -- 1.05 & 0.500 \\
    \hline
  \end{tabular*}
\end{table}

{\it Quantities to calculate}: The molar constant pressure heat capacity $c_p$ and
isothermal compressibility $\kappa$, both second order
partial derivatives of the Gibbs free energy, are the quantities of interest.
\begin{equation}
        \label{eq4}
        c_p=-\frac{k_B \beta^2}{N} \frac{\partial^2}{\partial \beta^2} (\beta G)
\end{equation}

\begin{equation}
        \label{eq5}
        \kappa =-\frac{1}{\beta \langle V \rangle} \frac{\partial^2}{\partial P^2} (\beta G)
\end{equation}
where $\beta=1/(k_B T)$, $G$
denotes the Gibbs free energy,  $P$ is the external pressure and $V$ is
the instantaneous volume of the system.
If there are anomalies in the behaviour of $c_p$ and $\kappa$ as a function of $P$ then a third order phase transition may indeed occur.
 
{\it Simulation set up}: We perform simulations in the isotherm-isobaric ensemble (NPT ensemble).
With this set up $c_P$ and $\kappa$ are calculated as: 
\begin{equation}
        \label{eq6}
        c_p=-\frac{1}{k_B T^2 N} \langle (H-\langle H \rangle)^2 \rangle
\end{equation}

\begin{equation}
        \label{eq7}
        \kappa=-\frac{1}{k_B T} \frac{\langle (V-\langle V \rangle)^2 \rangle}{\langle V \rangle}
\end{equation}
where $H$ is the enthalpy,  given by $H=\mathcal H
+PV$ with $\mathcal H $ being the Hamiltonian of the system.
Simulations are done at a certain temperature by fixing $\epsilon_B$, $\sigma_B$ and $\concen_B$ for a series of values of $P$. Along the isothermal line, the
second-order derivatives~\eqref{eq4} and \eqref{eq5} are then calculated.
If there were a third-order phase transition in these mixture systems,
then all second-order derivatives would present cusps (continuous but not
differentiable) at the transition point.

{\it Technical set up}: Each simulated system contains 4000 molecules interacting
via the L-J 12-6 potential in a periodic simulation box. The
conventional dimensionless unit system is employed. The unit of
length, energy, mass, and time are denoted by $\epsilon_A$, $\sigma_A$, $m$
and $\tau$, respectively. All quantities are written in the unitless
form by adding the superscript "$\ast$", e.g. $r^\ast=r/\sigma_A,
T^\ast=k_B T / \epsilon_A $, and $P^\ast=P\sigma_A^3/\epsilon_A$ . The MD time
step is $\Delta t^\ast=0.002$. Each simulation last for $1\times 10^8$ time
steps; the first $2.5 \times 10^7$ steps are then discarded to ensure that statistics is done in equilibrium. Every 100 time
steps the quantities of interest are sampled.  A Nose-Hoover
thermostat~\cite{nose1984molecular, hoover1985canonical}
and a Parrinello-Rahman barostat~\cite{parrinello1980crystal, parrinello1981polymorphic} are employed to generate an NPT
ensemble.
The cut-off radius in the simulations is chosen to be $r^\ast_c=8$.
The internal energy and the pressure
contribution from the molecules falling out of the cut-off radius are
included by the standard long-range correction.
All the simulation are done at the same temperature $T^\ast=1.36$;
such a value has been chosen based on results of our previous study
for the one-component system. $T^\ast=1.36$ was the lowest temperature
simulated, it was found that for higher temperatures the behaviour of
$c^\ast_p$ and $\kappa^\ast$ was much smoother and thus of no interest
for the existence of a phase transition. Since current systems are
progressive perturbation of the one-component system, one can expect
the same behaviour; indeed tests at higher temperature confirm this
trend (results are not shown). Moreover, sharp phase transitions happen only in the
ideal case of thermodynamical limit; in order to give a more solid ground to our conclusions
 we performed several tests considering systems of 8000 particles and of 16000 particles.
Results show that the behaviour of $c^\ast_p$ and $\kappa^\ast$ at the predicted point of discontinuity is not sensitive to the size of the system; moreover they are independent of the cut-off radius in the simulations.
The point of discontinuity represents the most delicate situation one can study for these systems, thus these additional tests suggest that the main conclusions of the work do not change in any sensible way by increasing the size of
the simulated system.  

\begin{figure}
  \centering
  \includegraphics[width=8cm]{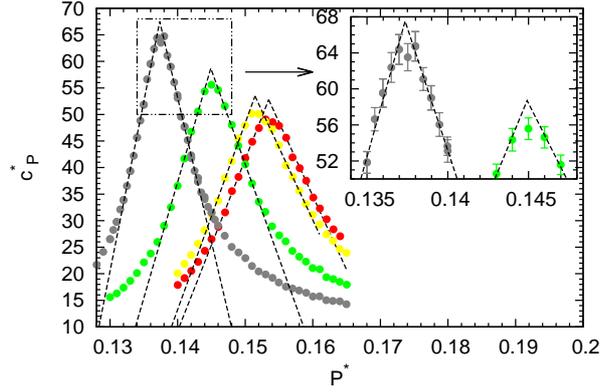}
  \caption{The molar constant pressure heat capacity as a function of
    pressure of four systems at $T^\ast=1.36$ and $\sigma^\ast_B=2.00$. In the
    main plot, four peaks from left to right correspond to
    $\concen_B=0.02, 0.01, 0.002, 0.00$, respectively. Error bars are
    not shown because they are smaller than the size of the
    dots. Two dashed straight lines are shown with each peak,
    presenting the linear regressing of data points on the higher and
    lower pressure branches of each peak, respectively. The inset
    shows the enlarged maximum of $c^\ast_p$, with error bars (indicating
    the confidence interval with 95\% confidence level) plotted on
    each data point. } 
\label{pic7}
\end{figure}

\begin{figure}
  \centering
  \includegraphics[width=8cm]{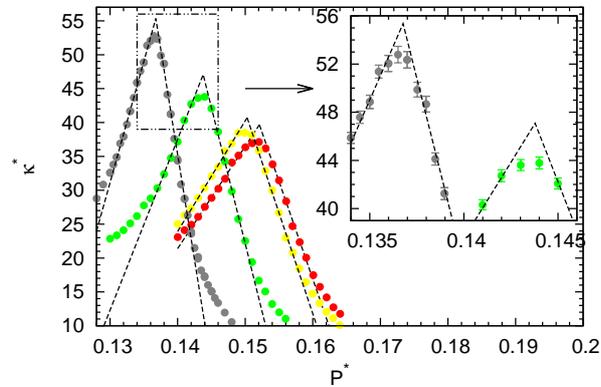}
  \caption{
    The same as Fig.~\ref{pic7}, except here the isothermal compressibility as a function of pressure is plotted.} 
\label{pic8}
\end{figure}

{\it Results for system (I)}:
Figs.~\ref{pic7} and \ref{pic8} show a clear trend: As $\concen_B$ becomes larger the positions of the
peaks shifts to the lower pressure side.  At the same time, the peaks
become sharper, which indicates the potential presence of a stronger singularity; however we 
have plotted two dashed lines that are linear regressions of
data along the low- and high- pressure branches of each peak.  If
there were a third-order phase transition, then the simulation data
would be consistent (within the statistical uncertainty) with the cusp
predicted by the intersection of the dashed lines.  It is not
difficult to draw the conclusion that there is no  indication whatsoever of a
third-order phase transition for $\concen_B = 0.002$, and 0.01, because
the heat capacity and compressibility, even considering the statistical uncertainty,
clearly do not follow the anomalous behaviour, which the predicted cusps
would suggest (see inset of Figs.~\ref{pic7} and \ref{pic8}).
The same conclusion could not be made for $\concen_B = 0.02$ by the plot of
the heat capacity  (Fig.~\ref{pic7}) due to the large statistical
error. Therefore, we have carried out an additional simulation at the crossing pressure, and
have proved that the cusp is clearly outside the error bar.
 
\begin{figure}
  \centering
  \includegraphics[width=8cm]{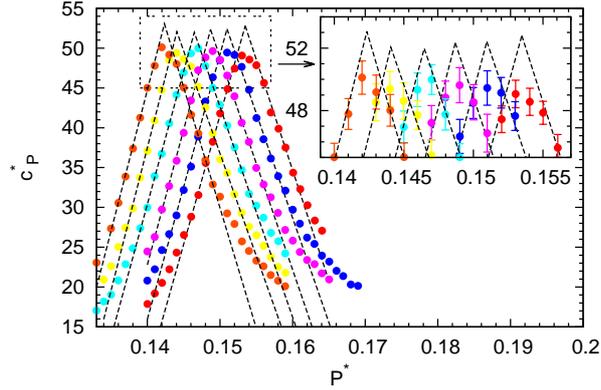}\\
  \caption{The molar constant pressure heat capacity $c^\ast_p$ as a function of
    pressure for six different systems at $T^\ast=1.36$ and $\concen_B=0.50$. In the main
    plot the six peaks from left to right correspond to $\sigma^\ast_B=1.05,
    1.04, 1.03, 1.02, 1.01, 1.00$, respectively. As before, the error bars are
    not shown because most of them are smaller than the size of the
    dots and the other technical details are the same of those of previous figures.} 
\label{pic1}
\end{figure}

\begin{figure}
  \centering
  \includegraphics[width=8cm]{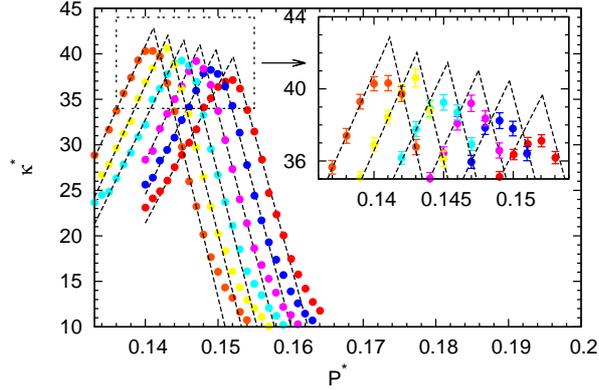}\\
  \caption{The same as Fig.~\ref{pic1}, except here the isothermal compressibility as a function of pressure is plotted.} 
\label{pic2}
\end{figure}

{\it Results for system (II):}
In Figs.\ref{pic1} and \ref{pic2} we plot the simulation
measurements of the mentioned second-order derivatives on the
isothermal lines at $T^\ast=1.36$ ; also in this case peaks of fluctuations appear along
each isothermal line together with the systematic shift of the
position of peaks towards the low pressure side.  However, in
contrast to system (I), the shapes of the peaks do not change with
respect to the increasing value $\sigma^\ast_B$, which suggests that the possibility of having a singularity 
at the peaks is independent of the system treated. 
However, the smoothness of curves around the peaks clearly shows that also in this case there is not evidence of a phase transition.

\begin{figure}
  \centering
  \includegraphics[width=8cm]{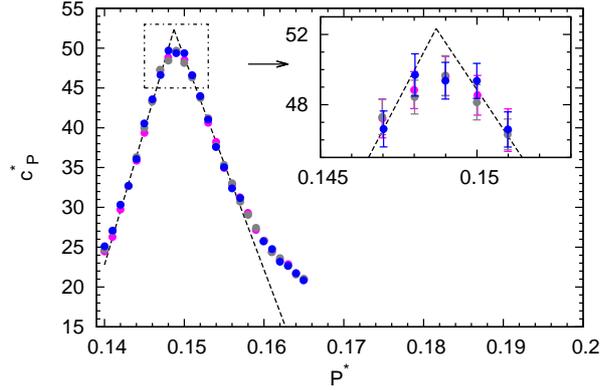}\\
  \caption{The molar constant pressure heat capacity as a function of
    pressure at $T^\ast=1.36$ for three different systems.
    The pink, gray and blue dots correspond
    to systems $\sigma^\ast_B=1.02,\ \concen_B=0.50$;
    $\sigma^\ast_B=1.05,\ \concen_B=0.20$;
    $\sigma^\ast_B = 1.04,\ \concen_B=0.25$, respectively. Error
    bars are not shown in the main plot, because of their negligible size. The
    inset shows the enlarged maximum of $c^\ast_p$. 
  } 
\label{pic5}
\end{figure}

\begin{figure}
  \centering
  \includegraphics[width=8cm]{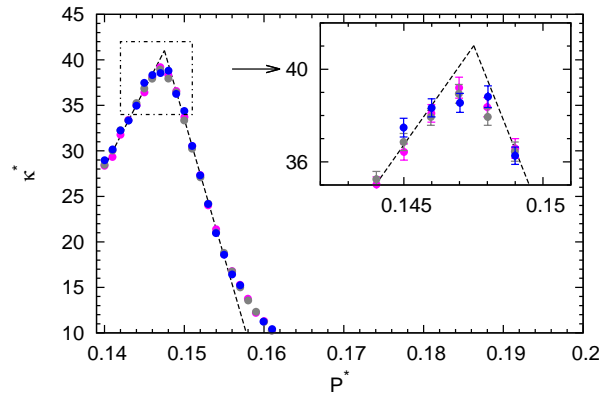}\\
  \caption{
    The same as Fig.~\ref{pic5}, except here the isothermal compressibility as a function of
    pressure is plotted.
  } 
\label{pic6}
\end{figure}

{\it Empirical law for the $\sigma^\ast_{B}$ v.s. $\concen_{B}$ behaviour:}
We have done further simulations showing that, chosen
$\sigma^\ast_{B}$, the distance (along the axis of the pressure) between
the peak of a mixture system and that of the one-component system
increase linearly by increasing $\concen_{B}$. This behaviour suggested
the question of what happens if we increase linearly $\concen_{B}$ and
decrease $\sigma^\ast_{B}$ or vice versa, can we obtain the same
thermodynamic behaviour from different systems by systematic
control of the parameters $\concen_{B}$ (concentration) and $\sigma^\ast_{B}$ (molecular size)?\\
Simulations show that the answer is positive, see Figs.~\ref{pic5} and
\ref{pic6}.
When we draw the results of $\sigma^\ast_B=1.05$, $\concen_B=0.20$;
$\sigma^\ast=1.02$, $\concen_B=0.50$ and $\sigma^\ast_{B}=1.04$, $\concen_B=0.25$ on the
same plot, we find that they coincide; interestingly the empirical law
linking the three systems is: $(\sigma^\ast_B-1)\times\concen_B=0.01$.  Despite at
this stage we do not have any deeper justification of this behaviour,
however, if it is physically realistic then it would suggest the
interesting opportunity of choosing either a specific fluid or a
specific concentration in order to obtain the same thermodynamic
behaviour.

{\it Discussion and Conclusions:} Supercritical fluids may indeed play
a crucial role for future technology, therefore the investigation of
their physical features is in this sense mandatory. The possibility of
the existence of a third order phase transition beyond the Andrew's
critical point is one of the most stimulating results of the last
years. We have investigated this possibility with extended MD
simulations of L-J mixtures, extending a previous study based on a
one-component system. Our results show a rather regular behaviour of
the fluid and suggest a negative response about the existence of such
a phase transition.  However, differently from experiments, done with
fluids with specific (chemical) molecular structure, our simulations
consider generic (chemically) unstructured molecules. In first
instance one may suppose that specific chemical structure may not play
a major role in the supercritical regime, as suggested before in this
work, however, this may not be true overall. Supposing that third order
phase transitions can be shown experimentally or numerically for
specific fluids, then our results implicitly show that the specific
molecular chemical structure may indeed be a key factor in the
thermodynamic behaviour in supercritical regime. If it was so, then
this would open exciting scenarios on how to chemically design liquids
with specific supercritical properties such as those related to the
third order phase transition. Clearly for liquids whose molecules
interact as L-J molecules, in simulation, phase transitions do not occur. In
conclusion, while showing that third order phase transition are in
general very unlikely on the basis of the current knowledge, this
paper, if proved wrong for specific systems, suggests that the
question about the relation between supercritical behaviour and
chemical nature of the fluid is a key issue of supercriticality.
In general, the negative response about the existence of
phase transition given by us does not represent an ultimate evidence of it; on the other hand, experimental and theoretical work, which provides a positive answer to the problem, does not give conclusive evidence although firmly claims the universality of the phase transition. In this context, our work must be considered as a pilot study for future theoretical investigations (hopefully at atomistic level) and should be carefully considered in the design of future experiments.

\section*{Acknowledgments}
This work was partially supported by the Deutsche Forschungsgemeinschaft (DFG) with the Heisenberg grant
provided to L.D.S (grant code DE 1140/5-1).
The work of P.Z. is supported by grants from National Natural Science Foundation of China (50930003, 21274005).
The results described in this paper were
obtained on the Deepcomp7000G of the Supercomputing
Center, Computer Network Information Center of Chinese
Academy of Sciences.


\end{document}